\documentclass[twocolumn, trackchanges]{aastex701} 
\definecolor{trackchange}{cmyk}{0,1.0,0.6,0.2}

\usepackage{multirow}

\usepackage{graphics,epsf}
\usepackage[utf8]{inputenc}
\usepackage{amsmath}                
\usepackage{amsfonts}               
\usepackage{amssymb}                
\usepackage{epsfig}                 
\usepackage{graphicx}               
\usepackage{float}
\usepackage{color}
\usepackage{multirow}               

\hypersetup{
    colorlinks=true,
    linkcolor=red,   
    urlcolor=cyan}

\usepackage[colorinlistoftodos]{todonotes}

\newcommand{\kms}{{~\rm km\; s^{-1}}}

\newcommand{\cm}{{~\rm cm}}
\newcommand{\km}{{~\rm km}}
\newcommand{\s}{{~\rm s}}

\newcommand{\g}{{~\rm g}}
\newcommand{\G}{{~\rm G}}
\newcommand{\K}{{~\rm K}}
\newcommand{\erg}{{~\rm erg}}
\newcommand{\yr}{{~\rm yr}}

\newcommand{\kpc}{{~\rm kpc}}

\begin{document}

\title{A pair of unequal jets imparted the kick velocity to the neutron star of Cassiopeia A}

\author[0000-0002-9444-9460]{Dmitry Shishkin}
\affiliation{Department of Physics, Technion - Israel Institute of Technology, Haifa, 3200003, Israel; s.dmitry@campus.technion.ac.il; akashi@technion.ac.il; soker@technion.ac.il}
\email{s.dmitry@campus.technion.ac.il}

\author[0000-0001-7233-6871]{Muhammad Akashi} 
\affiliation{Kinneret College on the Sea of Galilee, Samakh 15132, Israel}
\affiliation{Department of Physics, Technion - Israel Institute of Technology, Haifa, 3200003, Israel; 
s.dmitry@campus.technion.ac.il; akashi@technion.ac.il; soker@technion.ac.il}
\email{akashi@technion.ac.il}

\author[0000-0003-0375-8987]{Noam Soker}
\affiliation{Department of Physics, Technion - Israel Institute of Technology, Haifa, 3200003, Israel; 
s.dmitry@campus.technion.ac.il; akashi@technion.ac.il; soker@technion.ac.il}
\email{soker@physics.technion.ac.il}

\begin{abstract}
We analyze observations of the Cassiopeia A core-collapse supernova remnant (CCSNR) and identify a pair of opposite rings, attributing their formation to unequal-energy explosion jets that also imparted the neutron star (NS) its kick velocity via the kick-BEAP (kick by early asymmetrical pair of jets) mechanism. We replicate the formation of these ring structures, which we identify as circum-jet rings, with three-dimensional hydrodynamical simulations of two opposite explosion jets in which the northern jet is much more powerful, forming the larger northern ring known as the Crown. 
The line connecting the centers of the two opposite rings passes through the previously-established point-symmetric center of Cassiopeia A. The momentum difference between the two jets also explains the observed southwards NS kick velocity of $\simeq 400 \km \s^{-1}$. 
The kick-BEAP and point-symmetric morphology suggest powering and shaping by several pairs of jets during the explosion process, strengthening the jittering jets explosion mechanism for the Cassiopeia A CCSNR and core-collapse supernovae in general.
\end{abstract}


\section{Introduction} 
\label{sec:intro}

The Cassiopeia A supernova remnant (SNR) is one of the closest ($\approx3.5\kpc$, \citealt{Neumann_etal2024_CasAdist_LED}), and brightest (e.g., \citealt{2021MNRAS.502.5313S}) core-collapse supernova (CCSN) remnants (CCSNRs). Its young age (estimated at $\approx340 \yr$, \citealt{Fesenetal2006}) implies an early-phase, explosion-intrinsic, large-scale ejecta structure \citep{DeLaneyetal2010,Vink_etal2022_CasAshockDynamics}. Coupled with proximity, this makes Cassiopeia A a prime extended-source \citep[e.g.,][]{Reedetal1995,LamingHwang2003,Fesenetal2008} and compact central object \citep[CCO; e.g.,][]{Pavlovetal2000} scientific target.

The Cassiopeia A CCSNR (including its CCO) has been extensively used as an X-ray calibration source for Chandra and XMM \citep{Plucinsky_etal2025_CasAxrism}, and has historically served as a fundamental reference source for the radio flux-density scale \citep{Baars_etal1977_CasAradio}.
It is imaged both frequently, e.g., with Chandra \citep[e.g.,][]{HollandAshfordetal2024}, and with optical facilities \citep[e.g.,][]{FesenMilisavljevic2016,Milisavljevicetal2024}, and in depth, e.g., in X-ray \citep{Bambaetal2026}, and in IR \citep{Jungetal2026} - making it one of the best-studied CCSNRs.

The vast number of theoretical \citep[e.g.,][]{Young_etal2006_CasAprog,Orlandoetal2021} and observational \citep[e.g.,][and all the aforementioned]{DeLaneyetal2010,Grefenstetteetal2017,MilisavljevicFesen2015_SciBubble} works focused on the Cassiopeia A system emphasizes the need for thorough explanations for the explosion mechanism of CCSNe and the creation process of their compact remnants \citep[e.g.,][]{Young_etal2006_CasAprog, HoHeinke2009_CasAnsRemnant}.

Two heavily studied theoretical CCSN explosion mechanisms attempt to explain CCSNe by sourcing the explosion energy from the collapsing core.
The delayed neutrino-driven explosion mechanism \citep{BetheWilson1985_OGdelayedneutrino};
See \cite{Akaho2026,Andresenetal2026,Burrowsetal2026,Calvertetal2026, ChenCHetal2026, EggenbergerAndersenetal2026, LuoZhaKajino2026, Mezzacappa2026, Neopaneetal2026}, for some very recent papers on neutrino physics, mechanisms, simulations, and outcome predictions - within the framework of the neutrino-driven (neutrino-heating) mechanism.
The Jittering jets explosion mechanism (JJEM, \citealt{Soker2010, PapishSoker2011}); See \cite{Soker2026DustJets,Soker2026Failed,WangShishkinSoker2026,KlimovSoker2026,ShishkinSoker2026SNR0540,Soker2026RNAAS,AkashiSoker2026BG11,BraudoSoker2026Pipe} for some very recent papers on theoretical mechanisms, observational analysis and simulations - within the framework of jittering jets powered supernovae and their shaping of the explosion and subsequent remnant.

In rare cases, where the pre-collapse core is rapidly rotating, the neutron star (NS) launches a pair of fixed-axis jets to power an energetic CCSN  (\citealt{Shibataetal2025, Mannoetal2026, PanLi2026, Griffithsetal2026} for recent papers; a collapsar occurs in cases where a black hole is formed, e.g., \citealt{BoppGottlieb2025, Gottliebetal2025}). This is known as the magnetorotational mechanism.
In the neutrino-driven mechanism framework, the magnetorotational mechanism is a separate process that accounts for the most powerful CCSNe. In the JJEM, the magnetorotational mechanism is the extreme limit of a rapidly rotating pre-collapse core, resulting in small jittering of the jet axes around the pre-collapse angular momentum direction.\footnote{The magnetorotational mechanism is sometimes confused with the JJEM. However, the two mechanisms are significantly different (aside from both relying on jets to power the explosion): (1) The magnetorotational mechanism operates only rarely and produces energetic explosions, while the JJEM accounts for all CCSNe, including low-energy CCSNe.  (2) The angular momentum of the accretion disk around the newly born NS in the magnetorotational mechanism results from the rapidly rotating pre-collapse core, while in the JJEM, it results from the stochastic core convective motion. These differences lead to large differences in explosion outcomes. For a thorough discussion, see \cite{Soker2025G11}.} 

The neutrino-driven and jittering jets mechanisms share some common processes (e.g., \citealt{Soker2025Learning}). Both mechanisms require the pre-collapse core to have vigorous convection: to seed angular-momentum fluctuations in the JJEM and to set perturbations in the neutrino-driven mechanism.
In both mechanisms, a magnetar might add more energy to the ejecta. However, a magnetar operates only after the explosion. Also, in most cases with energetic magnetars, jets must drive the explosion because the neutrino-driven mechanism cannot supply the required explosion energy (e.g., \citealt{Kumar2025}). 
Neutrino heating operates in the JJEM, but it is not the primary driver of the explosion. Once jets propagate through the stalled shock, neutrino heating boosts their energy \citep{Soker2022nu}. 
    
At present, the main observable property that robustly distinguishes the two explosion mechanisms is the morphologies of CCSNRs (e.g., \citealt{Soker2024UnivReview, Soker2025Learning}). 
The JJEM predicts that a large fraction (but not all) CCSNe possess point-symmetric morphologies resulting from shaping by several pairs of jets with inclined axes, as most clearly seen in three-dimensional (3D) simulations of the JJEM (e.g., \citealt{Braudoetal2025, Braudoetal2026, AkashiSoker2026a, AkashiSoker2026BG11, BraudoSoker2026Pipe, BraudoSoker2026Precess}). While the JJEM naturally explains point-symmetric CCSNRs with pairs of opposite jets, the neutrino-driven and magnetorotational mechanisms do not explain point-symmetric structures in CCSNRs. 
The neutrino-driven mechanism can explain non-spherical geometries in CCSNRs, and even axial symmetry, as demonstrated in the explosion products of simulations in the framework of neutrino-driven explosions (e.g., \citealt{Wessonetal2026, Giudicietal2026}) - but not point-symmetric structures.

Starting with SNR 0540-69.3 (\citealt{Soker2022SNR0540, ShishkinSoker2026SNR0540}), studies identified about 20 CCSNRs with point-symmetric morphologies attributed to the JJEM; the last addition is SNR G7.7-3.7, where \cite{Luoetal2026G77} identified two pairs of opposite ears (which they termed blowouts), and \cite{Soker2026Long} attributed this morphology to the JJEM. 

\cite{BearSoker2025} used Cassiopeia A maps \citep{DeLaneyetal2010, Milisavljevicetal2024, Vinketal2024} to identify a rich point-symmetric morphology in Cassiopeia A, which comprises seven pairs of opposite (to the center) structural features, and three tentative pairs. They argue that this suggests shaping by at least two, and more likely more than four, pairs of opposite jets within the JJEM framework.

This study is motivated by the high-quality JWST observations of Cassiopeia A from the past 3 years, and the recent 3D hydrodynamical simulations of the JJEM that robustly show the formation of circumjet rings (e.g., \citealt{SokerAkashi2025, AkashiSoker2026a, AkashiSoker2026BG11}). 
We analyze Cassiopeia A within the JJEM framework, as we adopt the view that the neutrino-driven mechanism cannot account for its point-symmetric morphology; at best, it can explain only one axis. For example, \cite{Orlandoetal2021} suggested that the jet that formed the eastern jet-like structure was a post-explosion jet (for recent studies of Cassiopeia A's morphology in the framework of the neutrino-driven mechanism, see, e.g., \citealt{OrlandoJankaetal2025A, OrlandoJankaetal2025B, OrlandoJankaetal2026}, and \citealt{Bambaetal2026} for its iron distribution). 
In Section \ref{sec:Rings}, we will explore the two opposite rings in Cassiopeia A, mainly in JWST observations, which define a new symmetry axis.  
In Section \ref{sec:Jets}, we present 3D hydrodynamical simulations from which we estimate the power of the two jets that shaped the rings. In Section \ref{sec:Comparison}, we compare the simulated circumjet rings with three observed rings in Cassiopeia A.   

In Section \ref{sec:Kick}, we will apply the kick by an early asymmetrical pair of jets (kick-BEAP) to Cassiopeia A, and suggest that the energetic unequal pair of jets that shaped the unequal pair of rings also imparted to the NS its kick velocity. \cite{Bearetal2025Puppis} proposed the kick-BEAP and applied it to SNR Puppis A, where they identified three pairs of structural features indicating regular-energy pairs of jets, but also proposed one very energetic and early pair of jets with one jet much more powerful than the opposite one. 
\cite{Shishkinetal2025S147} applied the kick-BEAP to SNR S147, where they deduced that two misaligned pairs of unequal jets contributed to the NS kick. 

We summarize our study in Section \ref{sec:Summary} by strengthening the JJEM for Cassiopeia A and by further supporting the kick-BEAP mechanism as a common mechanism to impart natal kicks to newly born NSs.

\section{The Ring Structures of Cassiopeia A}
\label{sec:Rings}

Cassiopeia A's (hereafter, Cas A) bright, resolved structure makes for a popular scientific imaging target in both the X-ray \citep[e.g.,][]{Hwangetal2004} and (near-)visible bands, where bright knots can be used to define ejecta structures \citep{HammellFesen2008} - indicative of the remnant age and asymmetry \citep{Fesenetal2006}.
We focus on structures at the edge of the inner ejecta shell.

\begin{figure*}
\centering
\includegraphics[trim=0.0cm 0cm 0cm 0.0cm, width=\textwidth]{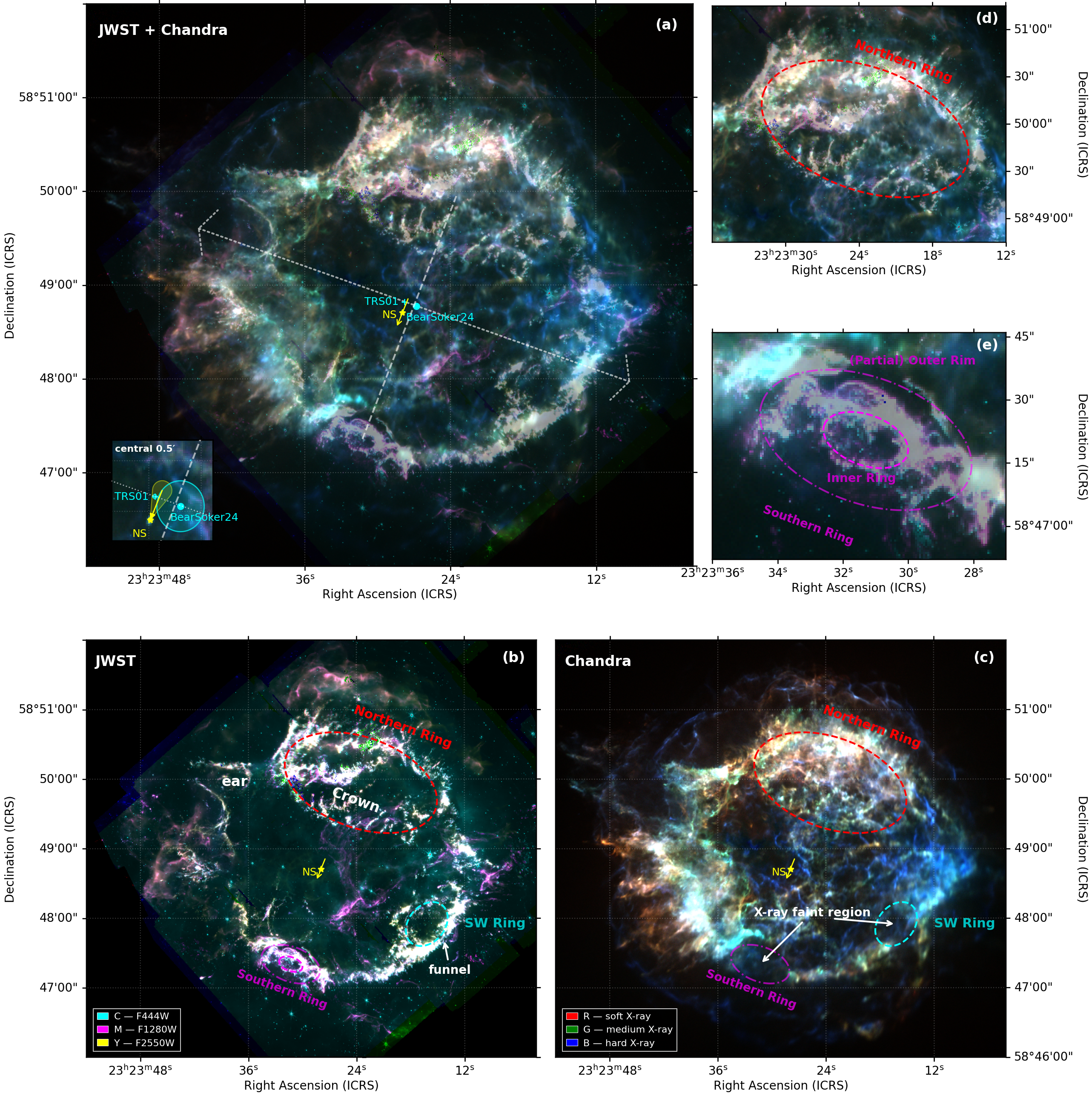}
\caption{Composite JWST and Chandra images of Cas A, with annotations denoting morphological features and symmetries. \\
\textbf{(a)} Stacked image of JWST band images (see panel (b)) and Chandra X-ray counts images (see panel (c)). The cyan markers show the expansion center of \cite{TFS2001_expCent} (marked with a $+$ sign) and the morphological center of \cite{BearSoker2025} (marked with a dot). The NS location is indicated by a yellow point. In the bottom-left inset, we show the central 0.5' region around the NS, with the NS trajectory as a yellow arrow and center confidences. \\
\textbf{(b)} Composite CMY image of F444W NIRCam and F1280W, F2550W MIRI filters. In addition to the center markings from panel (a), we mark the identified northern, southern, and SW rings with red, magenta, and cyan dashed ovals, respectively. 
\\
\textbf{(c)} Composite RGB image of three X-ray energy bands; $0.5-1.2\rm\ keV$ (soft), $1.2-2.0\rm\ keV$ (medium), $2.0-7.9\rm\ keV$ (hard). The northern ring is partially filled with X-ray emission; the southern and SW rings show gaps at the X-ray outer edge.
\\
\textbf{(d)} Zoom-in on the northern ring in the composite image (panel (a)).
\\
\textbf{(e)} Zoom-in on the southern ring in the composite image (panel (a)).
\\
}
\label{Fig:CasAobs}
\end{figure*}

As many of the ejecta morphological features are prominent in the visible and IR ranges, we show a composite JWST bands image in panel (b) of Figure~\ref{Fig:CasAobs}. We use the recent observations of \cite{Milisavljevicetal2024}, and select three representative bands to construct a cyan-magenta-yellow (CMY) image: NIRCam F444W, and MIRI F1280W, F2550W.\footnote{We use the level 3 data products as retrieved from the Mikulski Archive for Space Telescopes portal (program 1947 (PI: Milisavljevic), doi:\href{https://www.doi.org/10.17909/jtya-jj90}{10.17909/jtya-jj90}).}
We use a normalized log scale for the images to enhance features from all three different instrument bands. 

The northern part of Cas A contains a group of filament fingers stretching north, organized in an oval ring structure dubbed the "Crown" \citep[e.g.,][, see label in panel (b)]{MilisavljevicFesen2013}. We denote this feature with a red dashed ellipse and refer to it as the `northern ring' (see also \citealt{Satoetal2026b}).
On the opposite southern edge of the remnant bright shell is a small, full oval feature bright in the F1280W MIRI band. We mark this ring structure with a dashed magenta ellipse - the `southern ring'.
The southern ring has both a full inner ellipse, the inner rim (dashed magenta ellipse), and an outer partial ellipse (dotted-dashed magenta ellipse), defined by stray filaments and brighter emission in the edges of the SNR shell.

In the NE edge of the remnant is an extended outward feature that \cite{BearSoker2025} previously identified as an `ear' and an opposing feature on the SW that they mark as a funnel (as we mark on panel (b) of Figure~\ref{Fig:CasAobs}).
We now identify and mark with a dashed cyan ellipse a ring attached to the funnel and inner to the main CCSNR shell. We argue that this is a circumjet ring (Section \ref{sec:Comparison}).

To further define the ring features, we turn to X-ray count images. We use the 1Ms Chandra ACIS observations of \cite{Hwangetal2004} (Sequence number: 500459). We employ the standard CIAO \citep[][version 4.17.0, CALDB version 4.11.6]{CIAO2006} pipeline to reprocess, reproject, and recombine the 9 observations to form a combined tri-band image of the X-ray emission of Cas A. We show the resulting image in panel (c) of Figure~\ref{Fig:CasAobs}, where we use a similar normalized log scaling for the pixel intensity display as in the JWST data for visibility.

To visually connect features observed in the X-ray range with visible-band features, we stack the JWST image in CMY colors from panel (b) and Chandra image in red-green-blue (RGB) colors from panel (c), to produce the new composite image in panel (a) of Figure~\ref{Fig:CasAobs}.\footnote{Reprojection of the JWST data onto the Chandra data WCS was done using \textsc{reproject} from the Astropy package \citep{astropy:2013,astropy:2018,astropy:2022}.}

We note here that the Chandra observations were completed in the first half of 2004, whereas the JWST observations were carried out in late 2022. In light of the $340 \yr$ SNR age, this $18$-year difference makes for some small ($\approx5\%$) discrepancies in the combined X-ray-JWST image (panel (a)), but these are not substantial to our central claims.

In panel (d) of Figure~\ref{Fig:CasAobs} we show a zoom-in on the northern ring from panel (a), with the annotated ellipse.

The northern `Crown' feature's central region also exhibits strong X-ray emission in iron lines \citep[e.g.,][]{Bambaetal2026}, and progressively lighter elements dominate emission further from the feature center \citep[as seen in, e.g.,][]{Picquenot_etal2019_RedBlueShift}.
The ring-like distribution of ejecta material, with outward-extending fingers forming a crown-like morphology and enclosing bright iron emission, has been suggested to arise from stochastically generated, expanding nickel plumes that trigger Rayleigh–Taylor instabilities \citep{Orlandoetal2021, OrlandoJankaetal2025A}.
We focus on the spatial relationships among the individual rings and the Crown, with particular attention to their morphology.

Based on recent XRISM observations, \cite{Suzuki_etal2024_XRISM_CasA_Si} analyzed the velocity of intermediate mass elements (IME) ejecta. They find that along the line-of-sight (LOS) of the Crown feature, ejecta has bulk velocities of upwards of $5000 \kms$, with positive red-shifted velocities of $\sim1600 \kms$ and Doppler broadening of $\sim2000 \kms$.
\cite{Bambaetal2025CasA} measured the asymmetric expansion of Fe ejecta, similarly estimating Doppler velocities and Doppler broadening of $\sim1500 \kms$ in the same region.
The high Doppler broadening, especially compared to the thermal broadening of $\sim600\kms$, and specifically of the IME compared to the Fe, could suggest several ejecta components in that LOS \citep{Bambaetal2025CasA}.

\cite{Bambaetal2026} combined the XRISM observations with archival Chandra observations to produce a 3D expansion map of iron ejecta, where the Crown region has a net redshift.
Visible observations were previously used to produce detailed 3D reconstructions of Cas A; \cite{MilisavljevicFesen2013} created a full 3D map of the optically-emitting ejecta of Cas A. They identify a northern blue-shifted ring (in the same LOS as the Crown). This is consistent with the line images separating red- and blue-shifted components of \cite{Picquenot_etal2021_shiftedLines}.

In the south, several ring-like features composed of ejecta knots are located opposite the Crown and the smaller front-facing ring. One of them, nicknamed the `parenthesis', is suggested to lie opposite the Crown \citep{MilisavljevicFesen2015_SciBubble}, but is not identified as a counterpart to the Crown.
An interactive viewer made available by Dan Milisavljevic based on the data and reconstructions of \cite{MilisavljevicFesen2013,MilisavljevicFesen2015_SciBubble,DeLaneyetal2010} is available online\footnote{\href{https://www.physics.purdue.edu/kaboom/casa-webapp/}{https://www.physics.purdue.edu/kaboom/casa-webapp/}}.

We also identify several faint regions in the south and west of the bright X-ray shell seen in the counts image, which we denote in panel (c).
The outer rim of the southern ring coincides with an X-ray clearing, and the SW rings coincide with the funnel that \cite{BearSoker2025} identified. 
In panel (e) of Figure~\ref{Fig:CasAobs} we show a zoom-in on the southern ring as shown in panel (a), with the annotated ellipse over the inner and outer rims.

The ratio of the long to short axes of the northern and southern rings is $(a/b)= 1.75$. For circular 3D rings, this corresponds to a projection (ring-symmetry axis to line of sight) of $55^\circ$. Both rings have their short symmetry axis at $20^\circ$ to the north-south direction. 
We connect the centers of the two rings by the dashed white line on panel (a) of Figure~\ref{Fig:CasAobs}. We find that this line crosses the center of the point-symmetric structure that \cite{BearSoker2025} identified in Cassiopeia A (cyan dot on panel a of Figure~\ref{Fig:CasAobs}), but misses the center of the explosion that \cite{TFS2001_expCent} assessed (cyan '+'), and the NS present location (yellow asterisk) or projected origin (base of yellow arrow); this is better seen in the inset on the lower-left of panel (a).
The inset is the central 0.5 arcmin of the CCSNR. The yellow region is the confidence interval for the NS trajectory, which we propagated backward $320 \yr$ from the Chandra-observed NS position \citep{FesenPavlovSanwal2006_CasANS}, and taking the proper motion estimates of \cite{HollandAshfordetal2024} with their astrometry-corrected method confidences.

Our identification of the ring structures in Cassiopeia A is in part motivated by previous papers that explored the ejecta distribution \citep[e.g., ][]{MilisavljevicFesen2013,Kooetal2025}.
Another well-known structure in Cassiopeia A is the jet in the northeast and its counter-jet in the southwest (e.g., \citealt{Fesen2001_CasAoptical,Laming_etal2006_CasAjets, HammellFesen2008}).
It is documented in the IR \citep[e.g.,][]{Hinesetal2004}, the visible \citep[e.g.,][]{MilisavljevicFesen2013,Kooetal2023}, and X-ray \citep[e.g.,][]{Hwangetal2004,Satoetal2026}, and has motivated studies of the creation mechanism \citep[e.g.,][]{Ikedaetal2022, Zhanetal2022} and analysis methods \citep[e.g.,][]{Sakaietal2023,Sakaietal2024}.
We denote the general direction of this prominent NE-SW jet pair with a dotted, double-sided white arrow on panel (a) of Figure~\ref{Fig:CasAobs}. 
We adopt the definition of the NE/SW jets from \cite{HammellFesen2008}, who define two opposing conical regions from the directions of the optically emitting filaments. We connect these regions and shift the resulting line so that it passes through the remnant center(s).

The angle between the double-ring axis and the NE-SW jet-pair axis is $88^\circ$.

\begin{figure}
\centering
\includegraphics[trim=0.0cm 0cm 0cm 0.0cm, width=.45\textwidth]{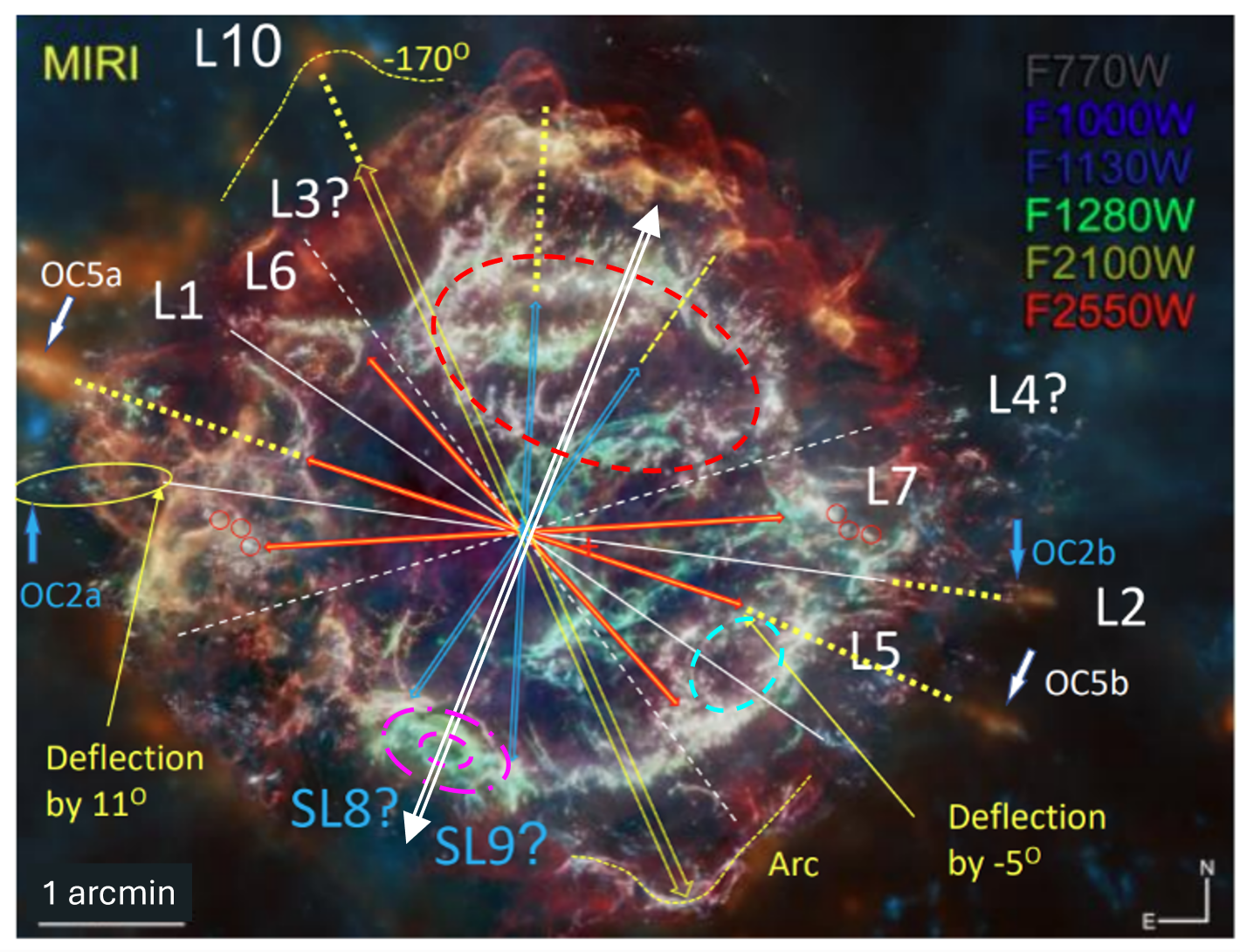}
\caption{
Figure from \cite{BearSoker2025}, showing the many point-symmetric features of Cas A. We add our newly denoted rings to their markings.
The SW ring coincides with the denoted L1 axis, matching the SW ring with the NE ear/nozzle feature. We add an additional new axis with a white double-sided arrow connecting the centers of the southern and northern rings of the new double-ring structure we identify.
}
\label{Fig:BearSoker2025}
\end{figure}

To place the double-ring system in the context of the point-symmetric morphology of Cassiopeia A, in Figure \ref{Fig:BearSoker2025} we add its symmetry axis (double-sided double-lined white arrow) to a figure from \cite{BearSoker2025}, where they marked 10 other symmetry axes connecting opposite structural features. 

The 11 symmetry axes now identified in the point-symmetric CCSNR Cassiopeia A do not suggest it was shaped by 11 pairs of jets.
We attribute each ear-ring and ring-ring pair, to a jet, but the clumps and filaments might be formed between jets \citep{Braudoetal2025, Braudoetal2026}. 
In any case, it seems that at least half this number, namely six or more, pairs of jets participated in the explosion of Cassiopeia A - fully compatible with the JJEM. 

The main result of this section is the identification of a double-ring structure that is part of the point-symmetric morphology of Cassiopeia A.  
As we show next, these rings are most likely circumjet rings.

\section{Simulating the ring formation} 
\label{sec:Jets}

\subsection{The numerical scheme} 
\label{subsec:Numerical}
\subsubsection{The initial spherical explosion} 
\label{subsubsec:Spherical}
We simulate the formation of bipolar circumjet rings resulting from the interaction of a pair of late jets with the expanding ejecta produced by an earlier spherical explosion. In our first paper in the series \citep{SokerAkashi2025}, we assumed that earlier jets exploded the core before the pair of late jets was launched, and we imposed a dense, fast-expanding shell in the core as an initial condition for the hydrodynamical simulations. In later papers \citep{AkashiSoker2026a, AkashiSoker2026BG11} we did not impose initial outflows but rather exploded the star by launching jets. 
In this study, we mimic the early explosion phase, which we assume includes many jet pairs, by launching a spherical wind. We base the assumption of many pairs of jets on the rich point-symmetric morphology of Cassiopeia A \citep{BearSoker2025}, as we present in Figure \ref{Fig:BearSoker2025}.
We inject an energy of $E_{\rm S}=2\times10^{51} \erg$ within $\Delta t_{\rm exp}=0.5\s$ starting at $t=0$. At $t=4 \s$, we launch a pair of unequal conical jets that propagate into the expanding ejecta and shape circumjet structures.

\subsubsection{The numerical grid} 
\label{subsubsec:Grid}
Due to limited numerical resources, we make several simplifications. (1) Although the jets' physical origin is expected to be at tens of km from the center, we launch them at thousands of km. (2) We neglect gravity. The initial explosion and the subsequent jets are sufficiently energetic that the ejecta in the interaction region are accelerated to velocities much larger than the local escape velocities. (3) We focus on the structures formed by the late jets in the outer parts of the ejecta and do not model the subsequent evolution of the innermost material.

We perform the three-dimensional hydrodynamical simulations using the Eulerian adaptive-mesh refinement (AMR) code \textsc{FLASH} v4.8 \citep{FryxellEtAl2000}, employing the unsplit hydrodynamics solver. The computational domain is a Cartesian box $(x,y,z)$ with
\begin{equation}
-1.8 \times 10^{10}\ \cm \le x,y,z \le 1.8 \times 10^{10}\ \cm,
\label{eq:size}
\end{equation}
corresponding to a cubic domain of side length
\begin{equation}
L = 3.6 \times 10^{10}\ \cm.
\label{eq:length}
\end{equation}

We apply outflow boundary conditions on the six cube faces. We use 7 refinement levels above the base grid, with one additional enforced level in the central injection region (maximum of 8 levels), giving an effective resolution of $2^{10}$ cells per dimension and a minimum cell size of
\begin{equation}
\Delta x_{\min} = 3.5 \times 10^{7}\ \mathrm{cm}.
\label{eq:deltax}
\end{equation}
Refinement is based on density and pressure gradients.

The stellar model follows our previous papers (see \citealt{AkashiSoker2026a} for the density profile), a stripped-envelope model of an $M_{\rm ZAMS} = 15 M_\odot$ star, adapted from \cite{Braudoetal2025}, and extending to a radius of $R_\ast=8\times10^{9}\ \cm = 0.115 R_\odot$. The innermost region of this stellar model is from \citet{PapishSoker2014Planar}, who fitted a post-bounce structure of a $15 M_\odot$ progenitor at $t \simeq 0.2\s$ after bounce \citep{Liebendorferetal2005}. 
Such stars require binary interaction to remove their envelope (e.g., \citealt{Gilkisetal2025}). Outside the stellar surface, the density drops sharply and connects to a lower-density profile that decreases approximately as expected for a stellar wind.

To avoid numerical difficulties near the origin, we impose an inner inert core of radius
\begin{equation}
R_{\rm core} = 4 \times 10^{8}\ \cm,
\label{eq:Rcore}
\end{equation}
within which we set the velocity to zero.

\subsubsection{Launching a late pair of jets}
\label{subsubsec:BipolarJets}

At $t=4\s$, after the initial spherical explosion has produced an expanding ejecta structure, we launch a pair of oppositely directed conical jets. Because of numerical limitations, we set the initial explosion phase and launch the jets at thousands of kilometers, where the dynamical time is longer than at the expected jet-launching zone of several tens of kilometers. This dictates a longer time delay from the initial explosion to the late pair of jets. In reality, this time can be $\simeq 1-2 \s$.  
We inject the jets within a time period of
\begin{equation}
\Delta t_{\rm jet}=1 \s, 
\label{eq:JetDt}
\end{equation}
and with the same initial velocity of 
\begin{equation}
v_{\rm jet}=6\times10^{9} \cm \s^{-1}. 
\label{eq:JetVelocity}
\end{equation}
The two opposite jets differ in their mass-outflow rates, hence energy, and half-opening angles.
The upper (northern) jet has a half-opening angle of $\alpha_{\rm up}=20^\circ$, and a mass outflow rate of $2\times10^{31}\ \g\ \s^{-1}$; hence, 
its total kinetic energy is
\begin{equation}
E_{\rm up}=3.6\times10^{50}\ \erg.
\label{eq:EnergyUpper}
\end{equation}
The down jet has a half-opening angle of  $\alpha_{\rm down}=5^\circ$, and an outflow mass rate of $5\times10^{29}\ \g$, so that its total kinetic energy is
\begin{equation}
E_{\rm down}=9.0\times10^{48}\ \erg.
\label{eq:EnergyLower}
\end{equation}
The two jets carry together $E_{\rm jets}
=3.69\times10^{50}\ \erg$, which is $ 15.6\%$ of the total explosion energy of $E_{\rm tot}
=E_{\rm S} + E_{\rm jets} = 2.369\times10^{51}\ \erg.$ 
\cite{Oralandoetal2016CasA} estimated the total kinetic energy of the ejecta of Cassiopeia A to be $\approx 2.3 \times 10^{51} \erg$. So the values we use here are plausible. If we include the binding energy of the ejecta (which we did not), the original jets must carry more energy to yield the final ejecta energy, so the jets' energy might be even larger.

During the jet-launching phase, the hydrodynamic quantities inside the two conical injection regions are overwritten ("reset") at every time step for the duration of the injection. Specifically, the jet density and velocity are imposed directly in these regions to match the prescribed jet properties. Thermal energy is not modified during injection and remains equal to that of the original ambient medium at the corresponding location. The thermal quantities remain consistent with the equation of state throughout the simulations.

The two opposite jets are deliberately assigned different properties to reproduce the general morphology of the double-ring structure of Cassiopeia A that we discussed in Section \ref{sec:Rings}. 

\subsection{Numerical results: circumjet rings} 
\label{subsec:Simulating}

We examine the evolution of the ejecta and the structures formed by the interaction of the jet pair with the expanding material produced by the initial spherical explosion. We focus on the evolution from $t=6 \s$ (1 second after the jets ceased) to $15 \s$, long after most material expanded beyond the original radius of the star. 

\begin{figure}
\hspace*{-1cm}
\centering
\includegraphics[trim=0.0cm 14.2cm 6.5cm 0.0cm, scale=0.54]{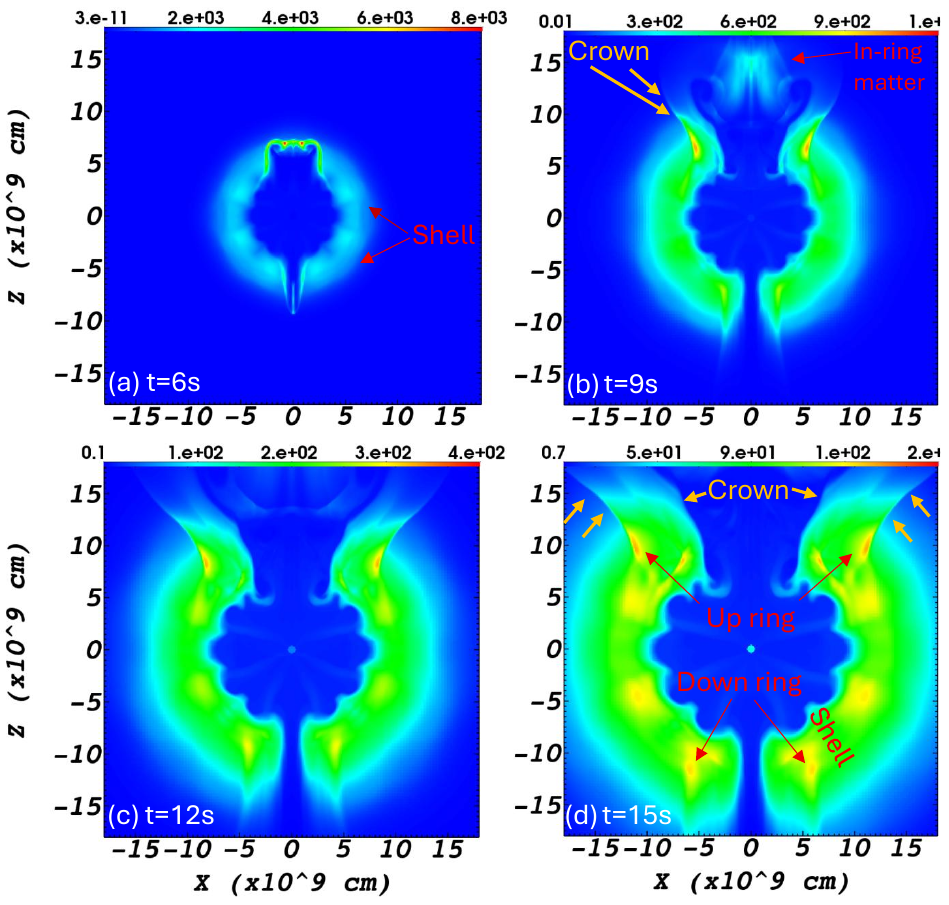}
\caption{Time evolution of the density in the meridional $xz$ plane at $t=6$, $9$, $12$, and $15 \s$. The color bar of each panel gives the density scale (differs between panels). We point to the structural features that form the brightest part of the up ring and down ring. The dense clumps (yellow spots) result from Rayleigh-Taylor instabilities, but their exact structure and distribution are dictated by the limited numerical resolution. 
}
\label{Fig:density_evolution}
\end{figure}

Figure \ref{Fig:density_evolution} presents density maps in the $xz$ meridional plane, showing the development of the large-scale structure that resulted from the interaction of the jets with the early ejecta; the $z$ axis is the jets' symmetry axis. The first (upper left) panel is at $t=6 \s$, close to the time the jets break out from the shell. Some density fluctuations result from limited numerical resolution, but others are real because the flow is unstable, as we discuss later. Although the up jet has 40 times more energy and momentum than the down-jet, its initial cross section is about 15 times larger, and the two jets break out at about the same time. Later panels show that the up jet opens a much larger hole in the shell. We point to the dense material that makes up the dense part of the ring that will also be bright in the numerical emission integral maps that mimic observations (the density maps have large-scale axial symmetry around the $z$-axis).   

The second panel of Figure \ref{Fig:density_evolution}, at $t=9 \s$, shows in-ring matter flowing at high velocity along the up jet axis (it is outside the grid in the two later panels). However, we simulate only the inner core. In reality, a massive hydrogen-rich envelope would stop this gas flow. If nucleosynthesis of $^{56}$Ni took place in this jet and the interaction it induced while colliding with dense core material, then this central gas will later be iron-rich. We suggest that this accounts for the strong iron concentration inside the northern ring in Cassiopeia A (Section \ref{sec:Rings}).  

\begin{figure}
\hspace*{-2cm}
\centering
\includegraphics[trim=1.4cm 18.9cm 5.2cm 0.0cm, scale=0.42]{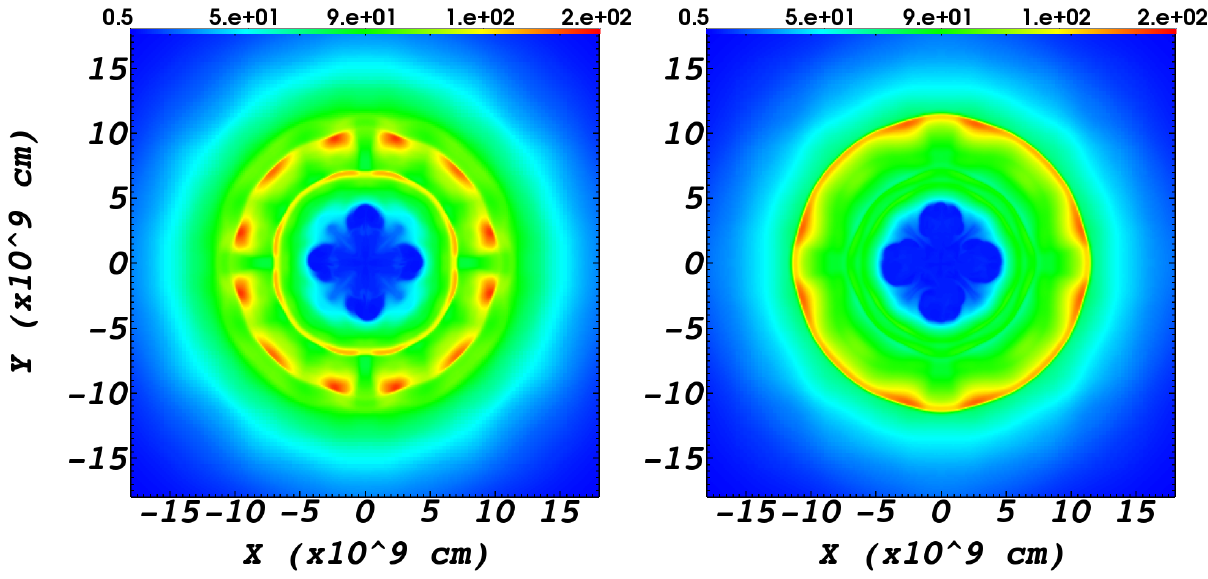}
\caption{Density maps in two planes perpendicular to the jets' axis and in the plane of the up ring: $z=8 \times 10^9 \cm$ (left) and $z=10 \times 10^{9} \cm$ (right). The left panel shows a rim inner to the main rim. This figure emphasizes the clumpy nature of the rings resulting from Rayleigh-Taylor instabilities. Densities are according to the color bar. }
\label{Fig:densXYUp}
\end{figure}

\begin{figure}
\hspace*{-3cm}
\centering
\includegraphics[trim=0.0cm 18.9cm 6.8cm 0.0cm, scale=0.40]{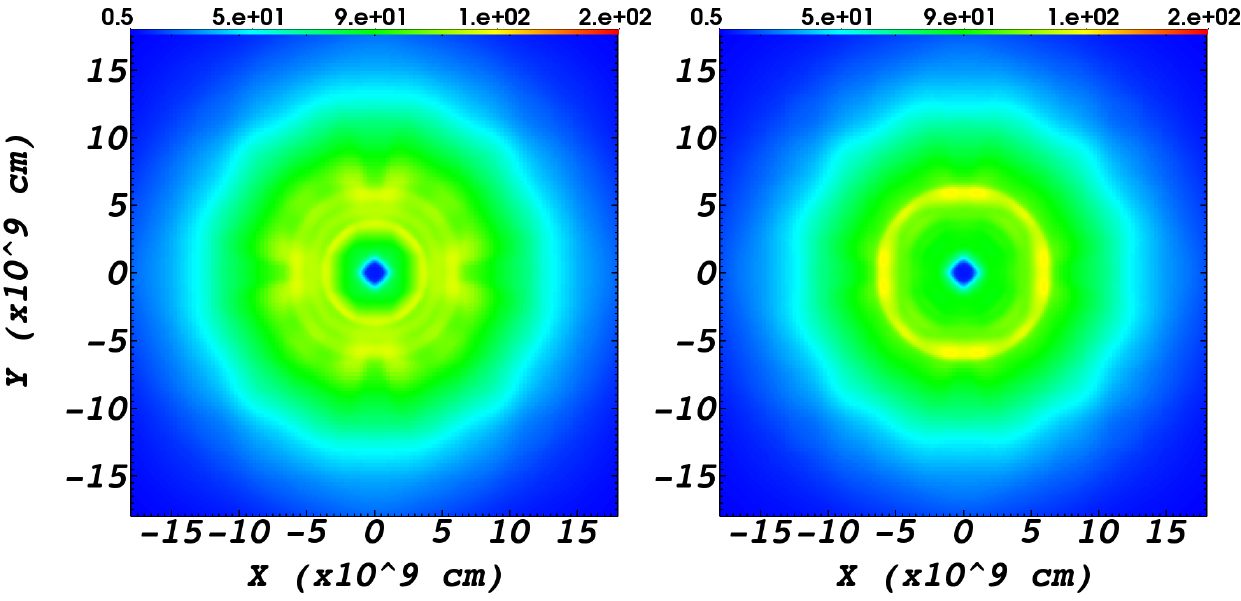}
\caption{Similar to Figure \ref{Fig:densXYUp}, but for two planes through the down ring: $z=-10.5 \times 10^9 \cm$ (left) and $z=-11.5 \times 10^{9} \cm$ (right). Note the bright inner rim in the left panel. }
\label{Fig:densXYdown}
\end{figure}

Late panels show that the upper ring surroundings include material extending outwards, the `Crown,' which we point at with orange arrows in Figure \ref{Fig:density_evolution}. The crown is not smooth but instead displays a filamentary structure in the meridional plane. The rings are also not smooth, but rather present clumpy structures. In Figure \ref{Fig:densXYUp} we show the density maps in the $z=8\times 10^9 \cm$ (left) and $z=10\times 10^9 \cm$ (right) planes that are perpendicular to the jets' axis, in the vicinity of the up ring. This figure shows that the up ring region consists of a fainter inner ring and the main ring (both in yellow-red in the figure), and that these structures are clumpy. In Figure \ref{Fig:densXYdown} we present two planes perpendicular to the jets' axis in the vicinity of the down ring. The plane $z=-10.5 \times 10^9 \cm$ shows an inner ring and an outer clumpy ring. The plane $z=-11.5 \times 10^9 \cm$ shows only the outer (main) ring, which we also point at in Figure \ref{Fig:density_evolution}. Generally, both the up and down rings show two rims. We also note that the plane of the up ring, the larger one, is closer to the center than the down ring (the smaller one). We see this also in Cassiopeia A: the northern and larger ring plane is closer to the center than the plane of the southern ring ($\approx73''$ and $\approx95''$, respectively).
  
\begin{figure}
\hspace*{1.5cm}
\centering
\includegraphics[trim=4.0cm -1.0cm 4cm 0.0cm, scale=0.60]{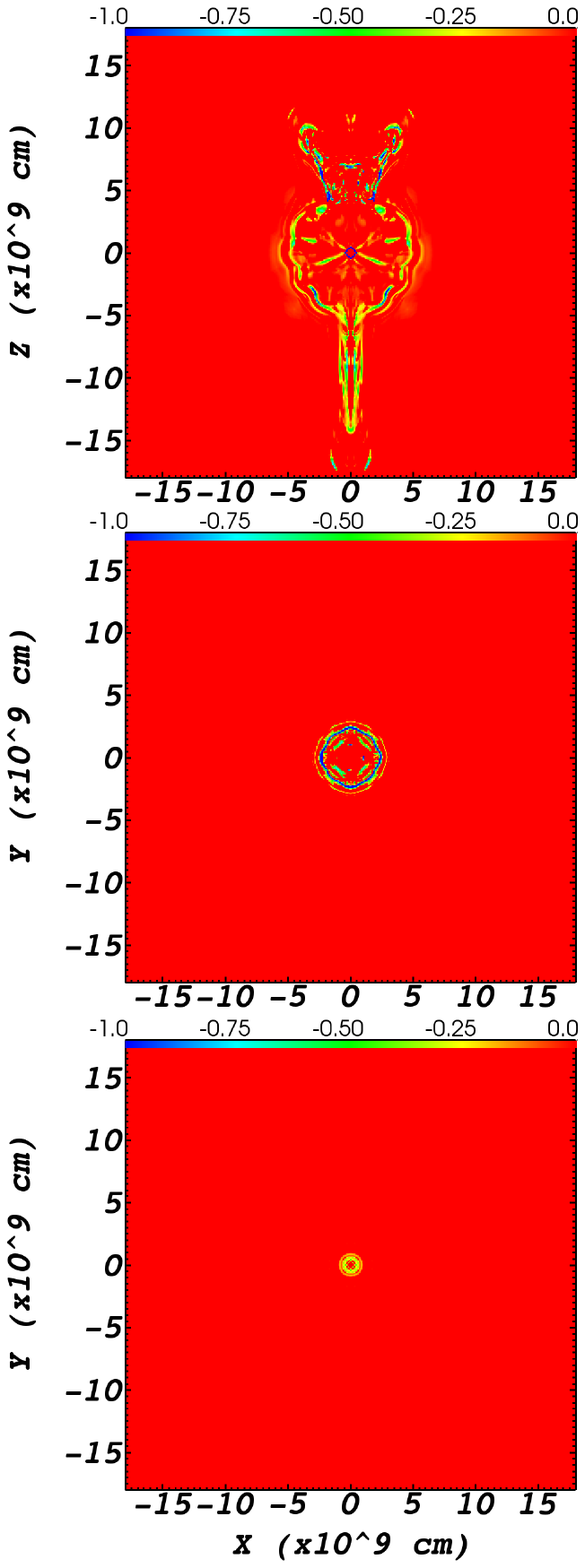}
\caption{Rayleigh-Taylor instability (RTI) growth rate ($f_{st}$; equation \ref{eq:rt}) maps at $t=7.5\s$ in three planes:  The meridional plane $y=0$ (top), a plane through the up ring $z=6.5\times 10^{9}\cm$ (middle), and through the down ring $z=-7\times 10^{9}\cm$ (bottom). Deep-red areas are RTI stable zones, while others are unstable with a typical growth time of $-1/f_{st}$. Color bar is from $-1 \s^{-1}$ (deep blue) to $\ge 0$ (deep red).}
\label{Fig:RTI}
\end{figure}

To check whether the formation of clumps in the ring results from Rayleigh-Taylor instabilities (RTIs), we follow the quantity 
\begin{equation}
f_{st} \equiv \frac{1}{\rho} \sqrt{\lvert \vec{\nabla}P \cdot \vec{\nabla}\rho \rvert} \ \text{sgn}(\vec{\nabla}P\cdot\vec{\nabla}\rho), 
\label{eq:rt}      
\end{equation}
where $\rho$ is the density, $P$ is the pressure, and ${\rm sgn} (\vec{\nabla}P\cdot\vec{\nabla}\rho)$ is the sign of the product of pressure and density gradients. If $f_{st}<0$ the region is unstable with a typical growth rate of $-f_{st}$, and a typical growth time of $-1/f_{st}$. 
In Figure \ref{Fig:RTI} we present maps of $f_{st}$ in three planes at $t=7.5 \s$: the meridional plane and two planes through the two rings.  
Figure \ref{Fig:RTI} shows that the interaction of the jets with the early ejecta is prone to RTIs, and that the growth time of the fastest-growing instabilities here is 1 second or shorter (blue regions); the green regions have a growth time of about 2 seconds. Limited numerical resolution and numerical viscosity suppress short-wavelength RTI modes; these are the modes that grow fastest. Our resolution is too coarse to resolve thin fingers as observed in the northern ring of Cassiopeia A (Section \ref{sec:Rings}). We conclude that RTI might explain the thin fingers in the northern ring of Cassiopeia A.   

\begin{figure*}
\centering
\includegraphics[width=\linewidth, trim=0.cm 9.0cm 0.cm 0cm, clip]{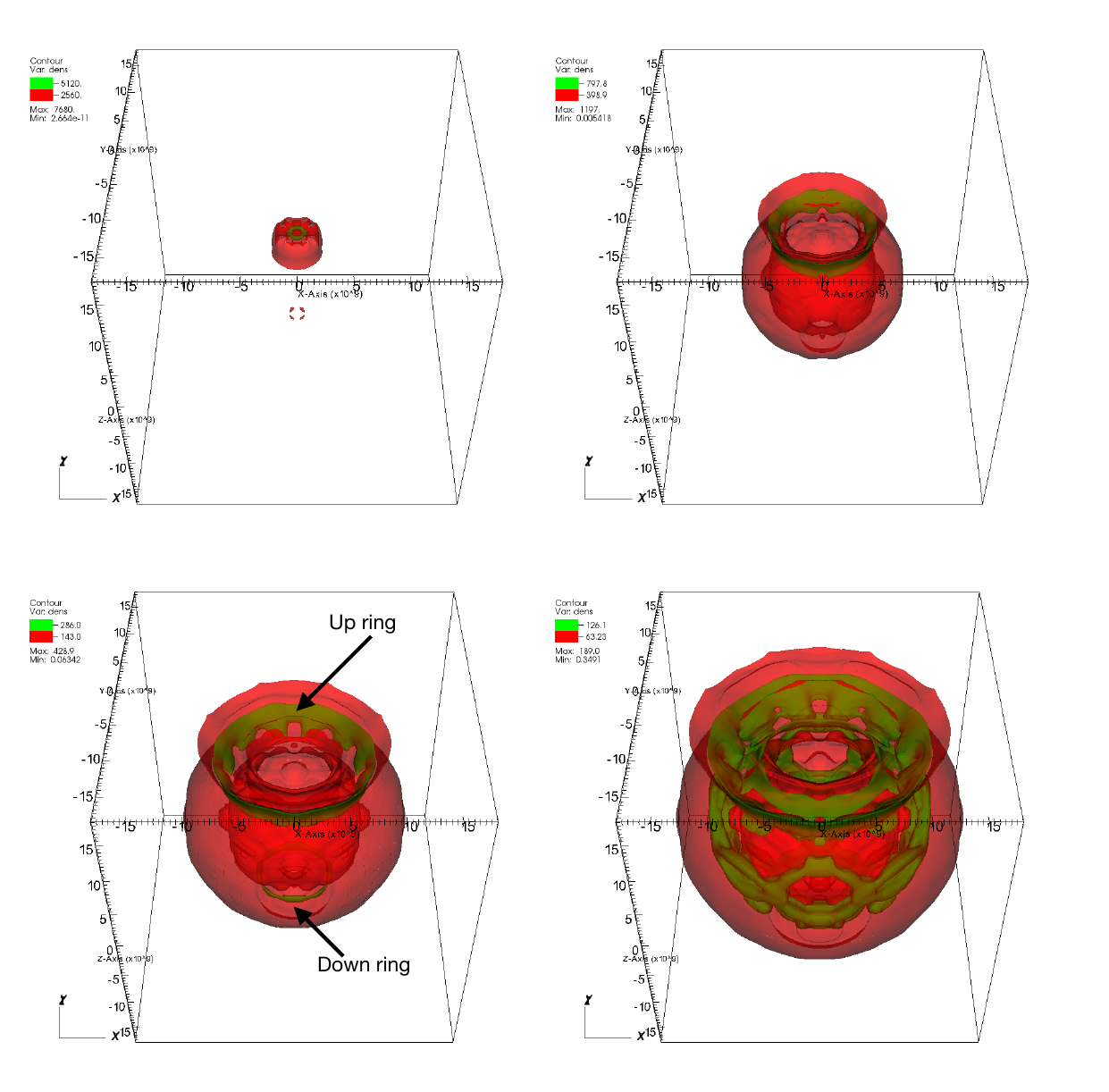}
\caption{Three-dimensional density structure at $t=6$, $9$, $12$, and $15 \s$, corresponding to the same times as in Figure \ref{Fig:density_evolution}. Each panel displays two semi-transparent equal-density surfaces (from upper left to lower right panel in $\g \cm^{-3}$): 
$(\rho_{\rm red},\rho_{\rm green})=(2560, 5120)$; $(399, 798)$; $(143, 286)$; and  $(63, 126)$. The green colors emphasize the double-ring structure and the clumpy and filamentary nature of the up ring, the deep red emphasizes the shell, and the uppermost red color away from the up ring is the Crown. }
\label{Fig:dens3D}
\end{figure*}

In Figure \ref{Fig:dens3D} we present a 3D density structure at the same four times as in Figure \ref{Fig:density_evolution}.  
These maps show two semi-transparent equidensity surfaces (the density values vary between panels). In these images, the up ring points toward us, and the down ring points away from us (the opposite of the rings in Cassiopeia A, but not important for this figure). The green color emphasizes the rings, and the deep red color the shell and regions around the rings. The last image at $t=15 \s$ (lower right) shows the following features. 
(1) A clear lower ring and upper ring (in green). (2) The ring structure is complicated: the up ring has a sub-ring inner to the main ring, while the down ring has a fainter ring further out (lower in the figure; in red). (3) The rings are not smooth, particularly the up ring, a manifestation of RTIs (the numerically limited resolution dictates the size and spacing of the instabilities). (4) The highest red regions in the panel show the Crown around and outside the up ring.   

\section{Comparison with Cassiopeia A} 
\label{sec:Comparison}

\begin{figure}
\centering
\includegraphics[trim=0.cm 14.4cm 5.0cm 0cm, scale=0.54]{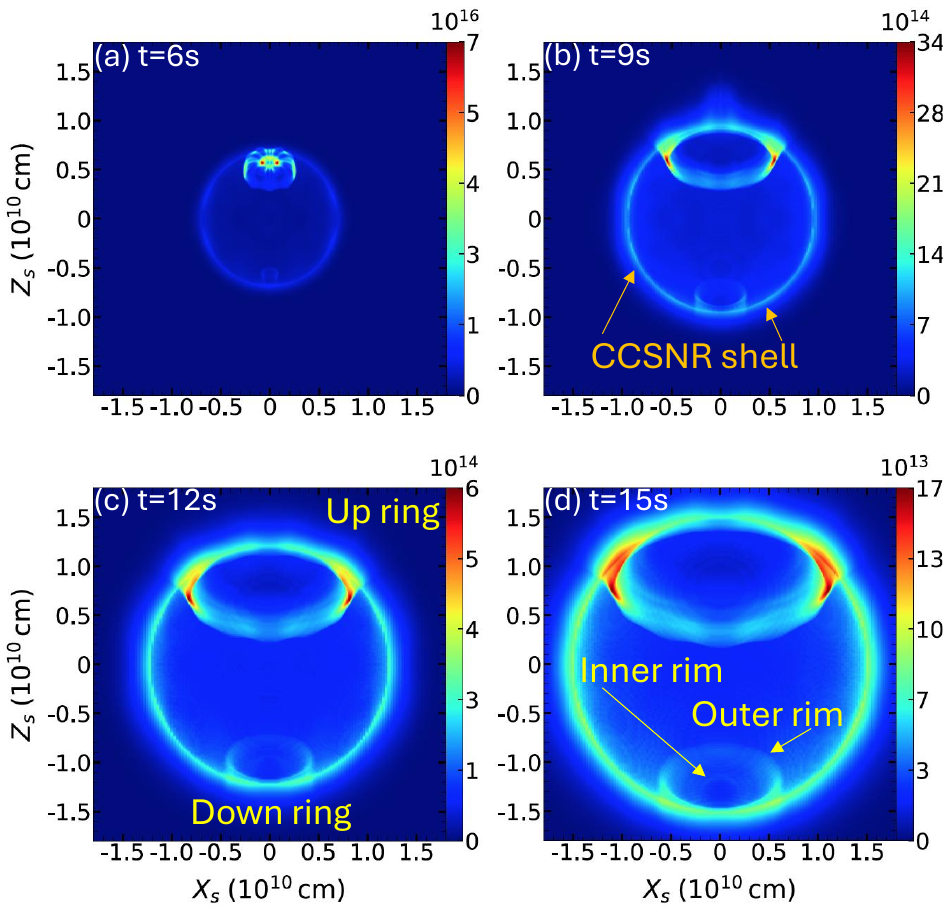}
\caption{Time evolution of the projected density squared (numerical emission integral $EI(X_s,Z_s)=\int \rho^2 dY_s$) at $t=6$, $9$, $12$, and $15\ \s$, viewed at an angle of $55^\circ$ between the $z$-axis of the simulation box and the line of sight. The color bars indicate the value of $EI(X_s,Z_s)$ in units of $\g^2 \cm^{-5}$. The projections highlight the dense structures of the circumjet rings produced by the interaction of the bipolar jets with the expanding ejecta.}
\label{Fig:projection}
\end{figure}

In Section \ref{subsec:Simulating} we described the properties of the pair of unequal rings that we reproduced with our simulation. Here we compare to the rings in Cassiopeia A, as described in Section~\ref{sec:Rings}. 
To mimic the observed structure, in Figure \ref{Fig:projection} we present the numerical emission integral $EI(X_s,Z_s)=\int \rho^2 dY_s$, where the integration is along the line of sight; we take the plane $(X_s,Z_s)$ to be the plane of the sky, and the coordinate $Y_s$ along the line of sight. Figure \ref{Fig:projection} presents $EI(X_s,Z_s)$ at an inclination of $55^\circ$, defined as the angle between the line of sight and the symmetry axis of the jets, i.e., the $z$-axis; this inclination reproduces the ratio of the long to short axes of the rings of $1.75$. 
The $X_s$-axis of the emission integral maps and the $x$-axis of the numerical grid coincide.  
Figure \ref{Fig:projection} shows how the rings become prominent with time. At $t=15 \s$ (lower right panel) the ratio of kinetic to thermal energy of the ejecta is $E_{\rm k}/E_{\rm th}=7$, implying that the ejecta is close to homologous expansion, which maintains the ring structure.    

We chose the energy of each jet to reproduce the approximate sizes of the rings of Cassiopeia A (although not exactly, as the boundary of the southern ring of Cassiopeia A is not well defined). We will discuss the energies of the jets in Section \ref{sec:Kick}. 

Although we simulated the formation of the north-south ring pair, we will also refer to the SW ring below. \cite{BearSoker2025} identified a funnel on that side (Figure \ref{Fig:CasAobs}) and did not refer to the ring; they marked a symmetry axis through the funnel that connects on the other side to a northeast (NE) ear. Here, we consider the SW ring the structural opposite of the NE ear, both along the symmetry axis L1 that \cite{BearSoker2025} identified.  

Such a pair of extended jet-like features on one side and a ring on the other is not unique to Cassiopeia A. 
The Crab Nebula has a well-studied northern jet-like feature (identified by \citealt{vandenBergh1970}; for the most recent study see \citealt{Dingetal2026}), very similar to the ear of Cassiopeia A but longer and narrower. Recently, \cite{Soker2026RNAAS} identified a ring opposite to the jet-like feature (or an elongated ear). The symmetry axis of the jet-like feature and the ring in the Crab Nebula is one of 9 symmetry axes that compose the rich point-symmetric morphology of the Crab Nebula; \cite{ShishkinSoker2025Crab} identified the other 8 symmetry axes and the point-symmetric morphology, and attributed the shaping to jittering jets in the JJEM framework. 

We compare the latest numerical emission integral image (at $t=15\s$, lower right panel of Figure~\ref{Fig:projection}) to the rings we identify in Section~\ref{sec:Rings}. 
In Table~\ref {tab:Rings}, 'Y' and 'N' refer to whether the ring possesses this property or not; 'Margin' means the ring marginally possesses this property. 
The properties we list in Table~\ref {tab:Rings} are as follows.
(1) The ring extends from the main CCSNR shell inward from the main CCSNR shell. 
(2) There is an inner rim clearly separated from the outer rim.
(3) The ends along the long axis of the rings are brighter, as clearly seen in the up ring (Figure \ref{Fig:projection}).
(4) There is an extension from the ring to larger distances from the center (a Crown).
(5) There are indications of prominent RTI. In the simulation, the down ring is unstable to the RTI, but the growth time is longer than for the up ring (as demonstrated in Figure~\ref{Fig:RTI}). So we mark `Margin'.   
\begin{table*}[!hbt]
\centering
\hspace{-1.5cm}
\begin{tabular}{l|ccc|cc}
\hline
& \multicolumn{3}{c|}{Observed rings} & \multicolumn{2}{c}{Simulated rings} \\
\textbf{} &
\textbf{North} &
\textbf{South} &
\textbf{SW} &
\textbf{Up} &
\textbf{Down} \\
\hline
Ring extends from the CCSNR shell inwards & Y & N & Y & \multicolumn{2}{c}{Both extend inwards} \\
Double Rim & N & Y & Margin & Margin & Y \\
Brightest on long-axis' edge & East & Y & N & Y & N \\
Crown & Y & \multicolumn{2}{c|}{No clear extended structure} &Y & N \\
Rayleigh-Taylor instabilities & Y & \multicolumn{2}{c|}{Non documented} & Y & Margin \\
\hline
\end{tabular}
\caption{Comparing several observed and simulated rings' morphological features. 
The observations refer to the three rings we mark on Figure \ref{Fig:CasAobs}, and the simulated rings to the lower-right images at $t=15 \s$ in Figures \ref{Fig:dens3D} and \ref{Fig:projection}. 
(1) For a large range of inclination angles away from $0^\circ$ and $90^\circ$, one side of the ring is projected at or near the main CCSNR shell. This side tends, therefore, to be brighter as it includes the emission from the shell. The other side is projected inside the main CCSNR shell (i.e., closer to the explosion site). The more complicated structure of the shell might prevent this, as in the south ring of Cassiopeia A. 
(2) The jet interaction with the CCSNR shell might form an extended ring (rather than a thin one), namely, with a large cross section. In some cases, there are two bright rims due to the compression of two annulus in the ring. 
(3) A purely inclined ring will be brighter on its edges along the long axis due to projection. This is not always the case when there is a more complicated structure due to the CCSNRs shell. 
(4) The crown is the extension of gas from the main ring outward away from the explosion site.
(5) In the north ring of Cassiopeia A this region suffered RTI and formed fingers, as we also obtain in the simulated up ring; in the down ring the growth rate is much slower. 
}
\label{tab:Rings}
\end{table*}

Table \ref{tab:Rings} shows that our simple simulation of a pair of jets launched into a spherical ejecta shell reproduced the large-scale structure of the double-ring system of Cassiopeia A, and some properties of the individual northern, southern, and SW rings.

We did not yet reproduce all individual properties. Other effects can play a role, such as more jets, varying jet properties, and a non-spherical shell or with sub-structure; since in the JJEM the early ejecta is formed by several pairs of jets, we do not expect it to be exactly spherical.
We consider our result to strongly suggest that a pair of unequal jets shaped the double ring structure of Cassiopeia A.  

\section{A kick-BEAP in Cassiopeia A} 
\label{sec:Kick}

The compact central object in Cassiopeia A \citep{Tananbaum1999IAUC} is an NS (e.g., \citealt{Pavlovetal2000, Zhaoetal2025}).
\cite{HollandAshfordetal2017} estimate the NS kick based on the optical-knot derived explosion center of \cite{TFS2001_expCent} as $\simeq 332\pm20 \km \s^{-1}$ in the direction of $10^\circ$ east of south.
\cite{HollandAshfordetal2024} use a 23-year X-ray observations baseline for a local rest-frame estimate of the NS velocity of $\simeq 433 \km \s^{-1}$ in the direction of $29^\circ$ east of south, as we mark with the yellow arrow in panels (a)-(c) of Figure \ref{Fig:CasAobs}. Unlike the previous method, this estimate does not include the proper motion of the progenitor (is not in relation to the SNR).
Overall, we take the kick velocity to be $v_{\rm k} \simeq 350-400 \km \s^{-1}$. 

The direction of the NS kick is the same as the symmetry axis of the double-ring structure we identified in Section \ref{sec:Rings}, and opposite to the direction of the large ring in the north. 
We therefore apply the kick-BEAP (kick by early asymmetrical pair), where a pair of highly unequal jets imparts momentum to the NS, hence contributing to its natal kick; the kick-BEAP was suggested by \cite{Bearetal2025Puppis} for the CCSNR Puppis A, and applied by \cite{Shishkinetal2025S147} to CCSNR S147. 
An unequal jet pair in the context of black hole formation consists of the relativistic limit of the jets described by the kick-BEAP mechanism.
\cite{LiGottleibMetzger2026} recently simulated a kick-BEAP-like mechanism in a collapsar, where a newly born black hole launches relativistic jets. They obtain wobbling jets, i.e., small changes in the jets' axes around the fixed angular momentum of the pre-collapse core (\citealt{Gottliebetal2022}). The unequal jets impart a kick to the black hole (their jet-asymmetry parameter $\alpha_{\rm jet}$ is equivalent to $f_{\rm j1}-f_{\rm j2}$ in \citealt{Bearetal2025Puppis}). 

An alternative explanation for NS kicks is the (gravitational) tug-boat mechanism (e.g., \citealt{Schecketal2004, Nordhausetal2010, Wongwathanaratetal2010, Nordhausetal2012, Janka2017} for theoretical frameworks and simulations), which has also been discussed in the context of the NS of Cassiopeia A (e.g., \citealt{Wongwathanaratetal2013kick, Wongwathanaratetal2017}).
More generally, the importance of rotation and possible spin-kick alignment has also been investigated (e.g., \citealt{YamasakiFoglizzo2008, LuBlackman2026}), alongside other proposed kick mechanisms (e.g., \citealt{Yaoetal2021, Xuetal2022}).
See also \cite{LambiasePoddar2025} for a recent review.

The two jets in our simulation carry together $E_{\rm jets}
=3.69\times10^{50} \erg$ (equations \ref{eq:EnergyUpper} and \ref{eq:EnergyLower};  Section~\ref{subsubsec:BipolarJets}).
The difference in the momenta of the two jets is $\Delta p_{\rm j} = 588 M_\odot \km \s^{-1}$. An NS of $1.4 M_\odot$ would acquire a velocity of $v_{\rm kick} \simeq 420 \km \s^{-1}$. 
We conclude that the kick-BEAP mechanism can explain the double-ring structure of Cassiopeia A and the kick velocity of its NS. 

The kick-BEAP mechanism can explain the spin-kick alignment observed in many pulsars (e.g., \citealt{Johnstonetal2005, Noutsosetal2012, BiryukovBeskin2025}), but not all. In many cases, but not all, it is accompanied by an energetic pair of jets, or at least one energetic jet aligned with the kick velocity. This is the case with Cassiopeia A, where the axis of the double ring structure is aligned with the kick velocity. It will not be aligned if there are two inclined kick-BEAP episodes, as \cite{Shishkinetal2025S147} suggested is the case with CCSNR S147. 

In addition, in many CCSNRs there is a symmetry axis at a large angle to the kick velocity, indicating a pair of jets with an axis at this angle to the kick velocity (e.g., \citealt{BearSoker2018kick, BearSoker2023RNAAS}). This is also the case with Cassiopeia A: the axis connecting the northeast prominent and long jet with its counter short jet in the southwest (the double-sided-dashed arrows in panel (a) of Figure \ref{Fig:CasAobs}), is at $88^\circ$ to the kick velocity. 
The explanation is that the NS launches the pair of jets at a large angle to the kick velocity after it acquired its kick velocity (e.g., \citealt{Soker2023postkick}). The kick velocity imparts angular momentum to the accreted gas that forms an accretion disk, which launches the jets; this angular momentum tends to be perpendicular to the kick velocity (e.g., \citealt{Soker202187Akick, Jankaetal2022spin}). 
The L1 axis, which connects the ear and the SW ring (panel b of Figure \ref{Fig:CasAobs} and Figure \ref{Fig:BearSoker2025}), is also at a large angle to the kick velocity. We proposed that the pair of jets that shaped this structure was also a post-kick pair of jets.

\section{Summary} 
\label{sec:Summary}

Cassiopeia A (Cas A) provides one of the clearest opportunities to deduce the dynamics of CCSN from the morphology of its remnant. Its ejecta exhibit a wealth of coherent knots, rings, filaments, and large-scale asymmetries whose geometries and kinematics retain information about both the explosion and the subsequent evolution. 

We identified a southern counterpart with signatures in both IR (JWST) and X-ray (Chandra) to the large northern structure, the Crown, of CCSNR Cassiopeia A (Section~\ref{sec:Rings}): a southern ring with two rims, as we marked in Figure \ref{Fig:CasAobs}. The two opposite rings share the same orientation and inclination, although the northern ring (the Crown) is much larger.
   
\cite{BearSoker2025} mark a symmetry line, L1, that connects an `ear' structure in the northeast with a funnel in the southwest. 
We identify SW ring (Figure \ref{Fig:CasAobs}) located to the inner side of the funnel as a potential counterpart to the ear (Figure \ref{Fig:BearSoker2025}). 
We consider the northern, southern, and SW ring as circumjet rings.

In Section \ref{sec:Jets}, we simulated the formation of the northern and southern rings in the JJEM framework as circumjet rings, with a late pair of jets during the explosion process shaping them as the jets interacted with the expanding core material that earlier energy deposition had exploded. We reproduced the relative ring sizes (Figures \ref{Fig:dens3D} and \ref{Fig:projection}) when the northern jet in our 3D hydrodynamical simulation is about 40 times as energetic as the southern one. The simulated circumjet rings and the three observed rings we studied share several features, as summarised in Table \ref{tab:Rings} and discussed in Section~\ref{sec:Comparison}. 
For example, we find the jet-shell interaction to be unstable to the RTI (Figure \ref{Fig:RTI}), a process that forms fingers; we attribute the filaments of the observed Crown to these instabilities. 
These similarities suggest that the northern and southern rings, as well as the SW ring, resulted from explosion jets, as predicted by the JJEM.

We connect the centers of the northern and southern rings (the double-sided double-lined arrow in Figure \ref{Fig:BearSoker2025}), and find that the line crosses at the center of the point-symmetric morphology that \cite{BearSoker2025} identified in Cassiopeia A. The ring pair we identify further enriches Cassiopeia A's point-symmetric morphology.

The unequal energies of the two jets that shaped the northern and southern rings would have also imparted a substantial kick to the CCO NS remnant of Cassiopeia A, i.e., the kick-BEAP mechanism (Section~\ref{sec:Kick}). With the parameters of the simulations we performed, the difference in jets' momenta yields, by momentum conservation, a kick velocity of $v_{\rm kick}\simeq420\kms$ for an NS mass of $1.4 M_\odot$, opposite to the direction of the stronger jet. The direction and velocity are very similar to the observed values (Section \ref{sec:Kick}). While the double north-south ring structure is along the kick velocity, the axis of the prominent east-west jet structures (dotted double-sided arrow in Figure \ref{Fig:CasAobs}) and the L1 axis of the ear-SW ring pair are almost perpendicular to the kick direction. In the JJEM, this implies that the NS launched these jet pairs after it acquired its kick velocity. 

The morphology of CCSNR Cassiopeia A attests to the kick-BEAP mechanism as the mechanism that imparted the kick to its NS, and a rich point-symmetric morphology supports the claim that several pairs of jets participated in the explosion process. 
These advance the JJEM for the Cassiopeia A CCSNR and further strengthen the claim that the JJEM is the primary explosion mechanism of CCSNe.

\section*{Acknowledgements}

A grant from the Pazy Foundation 2026 supported this research.
NS thanks the Charles Wolfson Academic Chair at the Technion for the support.

This work uses observations made with the NASA/ESA/CSA James Webb Space Telescope. The data were obtained from the Mikulski Archive for Space Telescopes at the Space Telescope Science Institute, which is operated by the Association of Universities for Research in Astronomy, Inc., under NASA contract NAS 5-03127 for JWST. These observations are associated with program \#1947.

The scientific results reported in this article are based in part on observations made by the Chandra X-ray Observatory with the following Observation IDs: 	
5196, 5320, 5319, 4634, 4635, 4636, 4637, 4638, 4639, and published previously in cited articles \citep{Hwangetal2004}.

\software{
NumPy \citep{harris2020array},
Matplotlib \citep{Hunter:2007},
Astropy \citep{astropy:2013, astropy:2018, astropy:2022},
CIAO \citep{CIAO2006},
FLASH \citep{FryxellEtAl2000},
VisIt \citep{Childs_High_Performance_Visualization--Enabling_2012}
}

\end{document}